\documentclass[12pt]{article}
\usepackage{epsfig}
\usepackage{cite}

\newcommand{\mysection}{\setcounter{equation}{0}\section}

\def\beq{\begin{equation}}
\def\eeq{\end{equation}}
\def\beqa{\begin{eqnarray}}
\def\eeqa{\end{eqnarray}}

\newlength{\dinwidth} \newlength{\dinmargin}
\begin{document}

\begin{center}
{\Large \bf Associated production of a top quark with a photon via anomalous couplings}
\end{center}
\vspace{2mm}
\begin{center}
{\large Matthew Forslund and Nikolaos Kidonakis}\\
\vspace{2mm}
{\it Department of Physics, Kennesaw State University,\\
Kennesaw, GA 30144, USA}
\end{center}
 
\begin{abstract}
We discuss the associated production of a top quark with a photon 
via anomalous $t$-$q$-$\gamma$ couplings, and present higher-order 
corrections from soft-gluon emission for the processes
$gq \rightarrow t\gamma$ at LHC energies. 
We perform soft-gluon resummation at next-to-leading-logarithm accuracy and derive approximate next-to-next-to-leading order (aNNLO) double-differential cross sections. 
We calculate at aNNLO the total $t\gamma$ production cross sections and the top-quark transverse-momentum and rapidity distributions for various LHC energies, and show that the soft-gluon corrections are large and dominant in these processes. 
\end{abstract}

\mysection{Introduction}
 
The top quark continues to play a central role at the Large Hadron Collider (LHC). It has the highest mass among all known elementary particles, and thus it has unique properties and is thought to be central to further understanding of the Higgs mechanism. The main production processes are top-antitop pair production and single-top production (see Ref. \cite{NKrev} for a review).

In models of physics beyond the Standard Model, it is possible to produce top quarks via anomalous top-quark couplings. One such process is the associated production of a top quark with a photon, see e.g. Refs. \cite{TY,NKAB,ZLLGZ,FG,TopC,DMWZ,DMZ,GYY}. While there are processes in the Standard Model for $t\gamma$ production with an additional quark in the final state, in models with anomalous top-quark couplings it is possible to produce a $t\gamma$ final state without extra particles.

We use an effective Lagrangian with an anomalous coupling, $\kappa_{tq\gamma}$, of a $t,q$ pair to a photon, with $q$ an up or charm quark,  
\begin{equation}
\Delta {\cal L}^{eff} =    \frac{1}{ \Lambda } \,
\kappa_{tq\gamma} \, e \, \bar t \, \sigma_{\mu\nu} \, q \, F^{\mu\nu}_\gamma + h.c.,
\label{Langrangian}
\end{equation}
where $\Lambda$ is an effective scale which 
we take to be equal to the top quark mass, $m$;  
$F^{\mu\nu}_\gamma$  is the photon field tensor; and 
$\sigma_{\mu \nu}=(i/2)(\gamma_{\mu}\gamma_{\nu}-\gamma_{\nu}\gamma_{\mu})$ 
where $\gamma_{\mu}$ are the Dirac matrices. These anomalous interactions
are flavor changing neutral currents. In the Standard Model 
such interactions are forbidden at tree level and highly suppressed.

The search for anomalous top-quark couplings is an active part of the LHC physics program \cite{CMS}. For a better theoretical input to the setting of experimental limits on anomalous couplings, it is important to include higher-order corrections, especially when the corrections are large. The next-to-leading order (NLO) corrections to $t\gamma$ production via anomalous couplings were calculated in Ref. \cite{ZLLGZ} and were found to be large. Therefore, for better theoretical control, next-to-next-to-leading order (NNLO) corrections need to be computed to determine their effect on the cross section.

Soft-gluon emission gives rise to a set of radiative corrections that are known to be large and dominant for final states with top quarks \cite{NKrev,NKAB,NKtop,NKts,NKtW,NKtZ}. For instance, in related processes such as $tW$ production \cite{NKtW} and $tZ$ production \cite{NKtZ}, the complete NLO results are very well approximated by the soft-gluon corrections at that order, and we find the same to be true for $t\gamma$ production. Below we denote as approximate NLO (aNLO) the cross section with first-order soft-gluon corrections; and as approximate NNLO (aNNLO) the cross section with second-order soft-gluon corrections. As we will see, the difference between the full NLO cross sections and the aNLO cross sections in $t\gamma$ production is negligible. The NNLO soft-gluon corrections make important contributions to the cross sections and differential distributions.

Near partonic threshold, where there is not much energy for additional radiation, contributions from soft-gluon emission are particularly important. Soft-gluon corrections appear in the form of ``plus'' distributions of a variable, $s_4$, that measures distance from threshold,
\beq
\left[\frac{\ln^{k}(s_4/m^2)}{s_4} \right]_+ \, , 
\label{lnplus}
\eeq
with $k\le 2n-1$ for the $n$th-order corrections, that arise from cancellations of 
infrared divergences between soft and virtual terms. 
Integrals of these perturbative plus distributions with nonperturbative parton distributions $\phi$ give
\beqa
\int_0^{s_{4 \, max}} ds_4 \, \phi(s_4) \left[\frac{\ln^k(s_4/m^2)}
{s_4}\right]_{+} &=&
\int_0^{s_{4\, max}} ds_4 \frac{\ln^k(s_4/m^2)}{s_4} [\phi(s_4) - \phi(0)]
\nonumber \\ &&
{}+\frac{1}{k+1} \ln^{k+1}\left(\frac{s_{4\, max}}{m^2}\right) \phi(0) \, .
\label{splus}
\eeqa
We will see that the soft-gluon corrections for $t\gamma$ production are significant. We also note that the effect of soft-gluon corrections is not dependent on the details of the models for top-quark anomalous couplings. In addition, we note that anomalous $t$-$q$-$g$ couplings can also contribute to $t\gamma$ production, and the formalism for the resummation of soft-gluon corrections is the same as for the $t$-$q$-$\gamma$ couplings.

In the next section we present our soft-gluon resummation formalism and its application to $t\gamma$ production. We derive formulas for the soft-gluon corrections at NLO and NNLO. In Section 3 we present numerical results at LHC energies for the total $t\gamma$ cross section and the top-quark transverse-momentum and rapidity distributions. We present our conclusions in Section 4.

\begin{figure}
\begin{center}
\includegraphics[width=81mm]{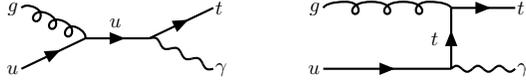}
\caption{Leading-order diagrams for the process $gu \rightarrow t\gamma$ via anomalous couplings.}
\label{lo-tgam}
\end{center}
\end{figure}

\mysection{Resummation for $gq \rightarrow t \gamma$}

We begin with the formalism for soft-gluon resummation, and 
develop it for $t\gamma$ production via anomalous $t$-$q$-$\gamma$
couplings. The relevant partonic processes are $gq \rightarrow t \gamma$, 
with $q$ denoting an up or charm quark. The leading-order diagrams 
for $gu \rightarrow t\gamma$ are shown in Fig. \ref{lo-tgam}, and identical 
ones apply to $gc \rightarrow t\gamma$.

We consider the partonic processes 
$g(p_g)+q(p_q) \rightarrow t(p_t)+\gamma(p_\gamma)$ 
and define the kinematical variables $s=(p_g+p_q)^2$,
$t=(p_g-p_t)^2$, $t_1=t-m^2$, $u=(p_q-p_t)^2$, and $u_1=u-m^2$, 
where $m$ is the top-quark mass, as before.
We also define the variable $s_4=s+t+u-m^2$, which vanishes 
at partonic threshold, where there is just enough
energy to produce the final $t\gamma$ state.

The cross section factorizes into functions that describe soft and collinear emission \cite{NKGS}. We take moments,   
${\hat \sigma}(N)=\int (ds_4/s) \;  e^{-N s_4/s} {\hat \sigma}(s_4)$, and write an expression for the double-differential cross section in $4-\epsilon$ dimensions,
\beq
\frac{d^2{\hat \sigma}_{gq \rightarrow t\gamma}(N,\epsilon)}{dt \, du}= 
H_{gq \rightarrow t\gamma} \left(\alpha_s(\mu)\right)\; S_{gq \rightarrow t\gamma} 
\left(\frac{m}{N \mu},\alpha_s(\mu) \right)\;
\prod_{i=g,q} J_i\left (N,\mu,\epsilon \right) 
\label{factsigma}
\eeq 
where $\alpha_s$ is the strong coupling, $\mu$ is the scale, $H_{gq \rightarrow t\gamma}$ is the hard-scattering function for the short-distance collision, 
$S_{gq \rightarrow t\gamma}$ is the soft-gluon function that describes  
noncollinear soft-gluon emission, 
and $J_i$ are jet functions that describe 
soft and collinear emission from the incoming (up or charm) quark and the incoming gluon.

The function $S_{gq \rightarrow t\gamma}$ obeys the renormalization group equation
\beq
\left(\mu \frac{\partial}{\partial \mu}
+\beta(g_s, \epsilon)\frac{\partial}{\partial g_s}\right)\,S_{gq \rightarrow t\gamma}
=-2 \, S_{gq \rightarrow t\gamma} \, \Gamma^S_{gq \rightarrow t\gamma}
\eeq
where $\beta(g_s, \epsilon)=-g_s \epsilon/2 + \beta(g_s)$ 
with $\beta(g_s)$ the QCD beta function, and 
$\Gamma^S_{gq \rightarrow t\gamma}$ is the soft anomalous dimension for 
the evolution of the soft function $S_{gq \rightarrow t\gamma}$.
The soft anomalous dimension $\Gamma^S_{gq \rightarrow t\gamma}$ is determined in dimensional regularization from the coefficients of the ultraviolet poles of the relevant loop diagrams \cite{NKtop,NKts,NKtW,NKAB,NKtZ,NKGS,NK2loop}.

The evolution of the functions in the factorized cross section results in the resummation of soft-gluon contributions. The resummed moment-space double-differential partonic cross section is given by 
\beqa
\frac{d^2{\hat{\sigma}}^{\rm resum}_{gq \rightarrow t\gamma}(N)}{dt \, du} &=&   
\exp\left[\sum_{i=g,q} E_i(N_i)\right]
H_{gq \rightarrow t\gamma}
\left(\alpha_s(\sqrt{s})\right) \;
S_{gq \rightarrow t\gamma}\left(\alpha_s(\sqrt{s}/{\tilde N'})
\right) 
\nonumber \\ && 
\times \exp \left[2\int_{\sqrt{s}}^{{\sqrt{s}}/{\tilde N'}} 
\frac{d\mu}{\mu}\; \Gamma^S_{gq \rightarrow t\gamma}
\left(\alpha_s(\mu)\right)\right]  \, ,
\label{resum}
\eeqa
where ${\tilde N'}=N(s/m^2) e^{\gamma_E}$, with $\gamma_E$ the Euler constant.
The first exponential in Eq. (\ref{resum}) resums soft and collinear 
corrections \cite{GS87,CT89} from the incoming up or charm quark and gluon,
and it can be found in \cite{NKtW}.

We write the perturbative series for the soft anomalous dimension for $gq \rightarrow t\gamma$ as $\Gamma^S_{gq \rightarrow t\gamma}=\sum_{n=1}^{\infty}(\alpha_s/\pi)^n
\Gamma^{S \, (n)}_{gq \rightarrow t\gamma}$.
For next-to-leading-logarithm (NLL) resummation we require  
the soft anomalous dimension at one loop. 
The one-loop expression in Feynman gauge is given by
\beq
\Gamma^{S\, (1)}_{gq \rightarrow t\gamma}=
C_F \left[\ln\left(\frac{-u_1}{m\sqrt{s}}\right)
-\frac{1}{2}\right] +\frac{C_A}{2} \ln\left(\frac{t_1}{u_1}\right)
\label{tgam1l}
\eeq
where $C_F=(N_c^2-1)/(2N_c)$ and $C_A=N_c$, 
with $N_c=3$ the number of colors.

The two-loop expression is given by 
\beq
\Gamma^{S\, (2)}_{gq \rightarrow t\gamma}=
\left[C_A\left(\frac{67}{36}-\frac{\zeta_2}{2}\right)
  -\frac{5}{18}n_f\right]  \Gamma^{S\, (1)}_{gq \rightarrow t\gamma}
+C_F C_A \frac{(1-\zeta_3)}{4}
\label{tgam2l}
\eeq
where $n_f=5$ is the number of light-quark flavors,
$\zeta_2=\pi^2/6$, and $\zeta_3=1.2020569\cdots$.

We expand the NLL resummed cross section, Eq. (\ref{resum}), to NNLO and then invert to momentum space, which gives us prescription-independent results for the soft-gluon corrections at NLO and NNLO. We will show in the next section that the NLO soft-gluon corrections are large and provide an excellent approximation of the complete corrections at that order. Furthermore, the NNLO soft-gluon corrections provide additional significant contributions. 

We provide analytical results with soft-gluon corrections for the double-differential partonic cross section $d^2{\hat\sigma^{(n)}}_{gq \rightarrow t \gamma}/(dt \, du)$ at $n$th order, with $n=0$, 1, and 2.
The LO cross section for $g q \rightarrow t\gamma$ is 
\beq
\frac{d^2{\hat\sigma^{(0)}}_{gq \rightarrow t \gamma}}{dt \, du}
=F^{\rm LO}_{gq \rightarrow t \gamma} \, \delta(s_4) \, ,
\label{LO}
\eeq
where 
\beq
F^{\rm LO}_{gq \rightarrow t \gamma}=
\frac{4 \pi \alpha \alpha_s \kappa_{tq\gamma}^2(m^2-s-t)}{3m^2s^3(m^2 - t)^2}
\left[m^6-m^4s-2st^2+m^2t(3s+t)\right] \, ,
\eeq
and $\alpha=e^2/(4\pi)$.

The NLO soft-gluon corrections for $g q \rightarrow t\gamma$ are given by
\beqa
\frac{d^2{\hat\sigma}^{(1)}_{gq\rightarrow t \gamma}}{dt \, du}
&=&F^{\rm LO}_{gq \rightarrow t \gamma} 
\frac{\alpha_s(\mu_R)}{\pi} \left\{
2 (C_F+C_A) \left[\frac{\ln(s_4/m^2)}{s_4}\right]_+ \right.
\nonumber \\ && \hspace{-23mm}
{}+\left[2 C_F \ln\left(\frac{u_1}{t}\right)-C_F
+C_A \ln\left(\frac{t_1}{u_1}\right)
+C_A \ln\left(\frac{s m^2}{u^2}\right)
-(C_F+C_A)\ln\left(\frac{\mu_F^2}{m^2}\right)\right] 
\left[\frac{1}{s_4}\right]_+
\nonumber \\ && \hspace{-23mm} \left.
{}+\left[\left(C_F \ln\left(\frac{-t}{m^2}\right)
+C_A \ln\left(\frac{-u}{m^2}\right)
-\frac{3}{4}C_F\right)\ln\left(\frac{\mu_F^2}{m^2}\right)
-\frac{\beta_0}{4}\ln\left(\frac{\mu_F^2}{\mu_R^2}\right)\right]  
\delta(s_4)\right\} \, ,
\label{NLOgqtgam}
\eeqa
where $\mu_R$ is the renormalization scale, $\mu_F$ is the factorization scale, 
and $\beta_0=(11C_A-2n_f)/3$ is the first term in the perturbative series for the $\beta$ function in QCD.

The NNLO soft-gluon corrections for $g q \rightarrow t\gamma$ are given by
\beqa
\frac{d^2{\hat\sigma}^{(2)}_{gq\rightarrow t \gamma}}{dt \, du}
&=&F^{\rm LO}_{gq \rightarrow t \gamma} 
\frac{\alpha_s^2(\mu_R)}{\pi^2} 2 (C_F+C_A) \left\{
(C_F+C_A) \left[\frac{\ln^3(s_4/m^2)}{s_4}\right]_+ \right.
\nonumber \\ && \hspace{-25mm}
{}+\frac{3}{2}\left[2 C_F \ln\left(\frac{u_1}{t}\right)-C_F
+C_A \ln\left(\frac{t_1}{u_1}\right)
+C_A \ln\left(\frac{s m^2}{u^2}\right)
-(C_F+C_A)\ln\left(\frac{\mu_F^2}{m^2}\right) -\frac{\beta_0}{6}\right]
\left[\frac{\ln^2(s_4/m^2)}{s_4}\right]_+
\nonumber \\ && \hspace{-25mm}
{}+\left[\left(3C_F \ln\left(\frac{-t}{m^2}\right)
-2C_F \ln\left(\frac{-u_1}{m^2}\right)+\frac{C_F}{4}
+3C_A \ln\left(\frac{-u}{m^2}\right)
+C_A \ln\left(\frac{u_1 m^2}{t_1 s}\right)-\frac{\beta_0}{4}\right)
\ln\left(\frac{\mu_F^2}{m^2}\right) \right.
\nonumber \\ && \hspace{-18mm} \left. 
{}+\frac{\beta_0}{2} \ln\left(\frac{\mu_R^2}{m^2}\right)
+\frac{1}{2}(C_F+C_A) \ln^2\left(\frac{\mu_F^2}{m^2}\right)\right]
\left[\frac{\ln(s_4/m^2)}{s_4}\right]_+
\nonumber \\ && \hspace{-25mm} \left.
{}+\left[-\frac{\beta_0}{4}\ln\left(\frac{\mu_F^2}{m^2}\right)\ln\left(\frac{\mu_R^2}{m^2}\right) 
+\left(\frac{3\beta_0}{16}+\frac{3}{8}C_F-\frac{C_F}{2}\ln\left(\frac{-t}{m^2}\right)
-\frac{C_A}{2}\ln\left(\frac{-u}{m^2}\right)\right)\ln^2\left(\frac{\mu_F^2}{m^2}\right) \right] \left[\frac{1}{s_4}\right]_+ \right\} \, .
\nonumber \\ 
\label{NNLOgqtgam}
\eeqa

\mysection{Results for $gu\rightarrow t\gamma$ and $gc\rightarrow t\gamma$ at the LHC}

Numerical results for $t\gamma$ production are now presented for $pp$ collisions at LHC energies via anomalous $t$-$u$-$\gamma$ and $t$-$c$-$\gamma$ couplings. MMHT2014 \cite{MMHT} NNLO parton distribution functions (pdfs) are used for all numerical results; however, the results are practically unchanged if CT14 \cite{CT14} or NNPDF \cite{NNPDF} pdfs are chosen instead. 

We first compare our results to the NLO calculation in Ref. \cite{ZLLGZ}. We note that our aNLO cross sections are very similar to the NLO cross sections in \cite{ZLLGZ}. The NLO/LO ratio at the LHC in \cite{ZLLGZ} is 1.36 while our aNLO/LO ratio (with the same parameters as used in \cite{ZLLGZ}) is 1.35. So our aNLO results are in remarkable agreement, to better than one percent, with the complete NLO calculation. This agreement conclusively demonstrates that the overwhelming majority of the NLO corrections come from soft-gluon corrections. It also shows the power and usefulness of soft-gluon resummation methods, and is in line with similar results for related processes \cite{NKtop,NKts,NKtW,NKtZ}.

\begin{figure}
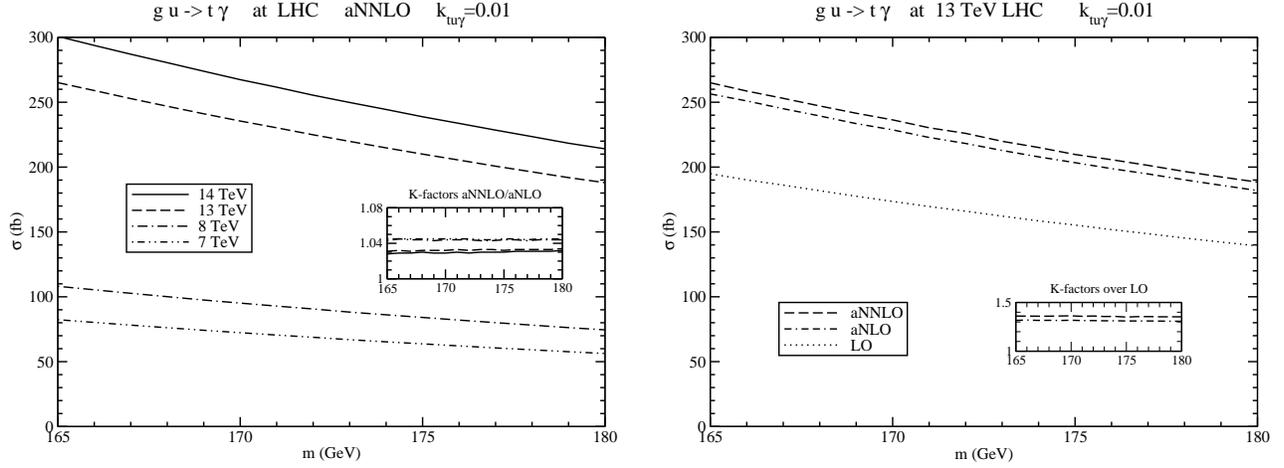

\begin{center}
\includegraphics[width=81mm]{gutgamNNLOm.eps}
\hspace{3mm}
\includegraphics[width=81mm]{gutgam13m.eps}
\caption{Total cross sections for the $gu\rightarrow t\gamma$ process with anomalous $t$-$u$-$\gamma$ coupling for LHC energies. The left plot shows aNNLO results for 7, 8, 13, and 14 TeV energies. The right plot shows LO, aNLO, and aNNLO results for 13 TeV energy. The inset plots display $K$-factors discussed in the text.}
\label{gutgam}
\end{center}
\end{figure}

We begin by presenting our cross sections at LHC energies for the process $gu \rightarrow t\gamma$. Based on recent limits \cite{CMS}, we use a value for the anomalous coupling of $\kappa_{tu\gamma}=0.01$ throughout. Fig. \ref{gutgam} displays total cross sections for $gu \rightarrow t\gamma$ as functions of top-quark mass ranging from 165 GeV to 180 GeV. We set the factorization and renormalization scales equal to the top-quark mass. 

The left plot in Fig. \ref{gutgam} displays aNNLO total cross sections calculated at LHC energies of 7, 8, 13, and 14 TeV. The plot's inset shows the $K$-factors relative to aNLO; that is, the aNNLO/aNLO cross section ratios for each LHC energy. We find that the aNNLO corrections increase the aNLO cross section by about 3.0\% at 14 TeV, 3.3\% at 13 TeV, 4.3\% at 8 TeV, and 4.4\% at 7 TeV. As expected, the effect of the corrections increases as the energy is lowered, since we are approaching the partonic threshold, where the soft-gluon corrections contribute more.

The plot on the right side of Fig. \ref{gutgam} shows the cross sections at each order for $gu  \rightarrow t\gamma$ at 13 TeV energy. The LO, aNLO, and aNNLO results are displayed at the same energy to demonstrate the effect of the perturbative soft-gluon corrections. It is immediately evident that the NLO soft-gluon corrections increase the LO result by a significant amount. The NNLO corrections also contribute significantly to the cross section, though less so than the NLO corrections, and must be included to make more accurate theoretical predictions.

The inset on the right plot of Fig. \ref{gutgam} displays $K$-factors, this time relative to LO; that is, the aNNLO/LO and aNLO/LO cross section ratios. The aNLO/LO $K$-factor is 1.31, indicating a 31\% enhancement of the LO cross section by the aNLO corrections. The aNNLO/LO $K$-factor is 1.36, indicating a further enhancement of the cross section by the NNLO soft-gluon corrections, for a 36\% total increase of the LO cross section.

\begin{figure}
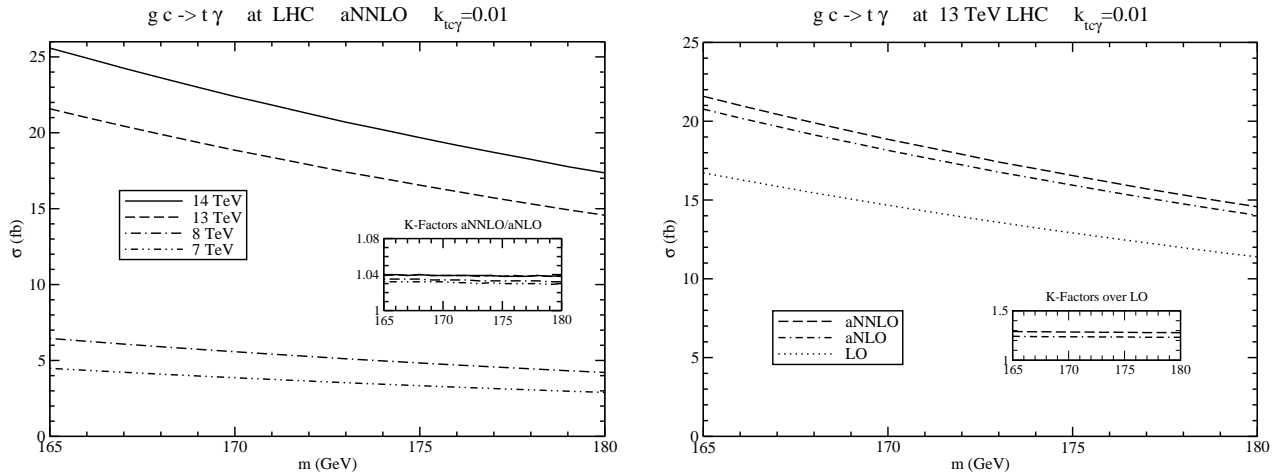

\begin{center}
\includegraphics[width=81mm]{gctgamNNLOm.eps}
\hspace{3mm}
\includegraphics[width=81mm]{gctgam13m.eps}
\caption{Total cross sections for the $gc\rightarrow t\gamma$ process with anomalous $t$-$c$-$\gamma$ coupling for LHC energies. The left plot shows aNNLO results for 7, 8, 13, and 14 TeV energies. The right plot shows LO, aNLO, and aNNLO results for 13 TeV energy. The inset plots display $K$-factors discussed in the text.}
\label{gctgam}
\end{center}
\end{figure}

We now move on to the process $gc \rightarrow t\gamma$. A value of $\kappa_{tc\gamma}=0.01$ is chosen for the anomalous coupling, by the same reasoning as for $\kappa_{tu\gamma}$. Fig. \ref{gctgam} displays total cross sections for $gc \rightarrow t\gamma$ as functions of top-quark mass over the same range of 165 GeV to 180 GeV. The cross sections for this process are an order of magnitude smaller than for $gu \rightarrow t\gamma$. The factorization and renormalization scales are again set equal to the top-quark mass. The left plot displays aNNLO total cross sections calculated at the same LHC energies of 7, 8, 13, and 14 TeV, as for $gu \rightarrow t\gamma$. The inset in the left plot again shows $K$-factors relative to the aNLO cross section. They are again between 3-4\%, very similar to those found for $gu \rightarrow t\gamma$. 

The right plot of Fig. \ref{gctgam} shows the cross sections at each order for $gc \rightarrow t\gamma$ at 13 TeV energy, similar to before. The $K$-factors with respect to LO are displayed on the inset. The aNLO/LO $K$-factor is about 1.24, and the aNNLO/LO $K$-factor is about 1.28. While not as large as with the process $gu \rightarrow t\gamma$, these are still significant enhancements over LO.

Since the higher-order corrections are significant, their effect on setting limits on the $t$-$q$-$\gamma$ anomalous couplings is important. For the $t$-$u$-$\gamma$ coupling at 13 TeV energy, the NLO corrections decrease the limit on the coupling by 14\% relative to LO, and with the addition of the aNNLO corrections the decrease is 17\% relative to LO. For the $t$-$c$-$\gamma$ coupling at 13 TeV energy, the NLO corrections decrease the limit on the coupling by 11\% relative to LO, and with the addition of the aNNLO corrections the limit is decreased by 13\% relative to LO. 

\begin{figure}
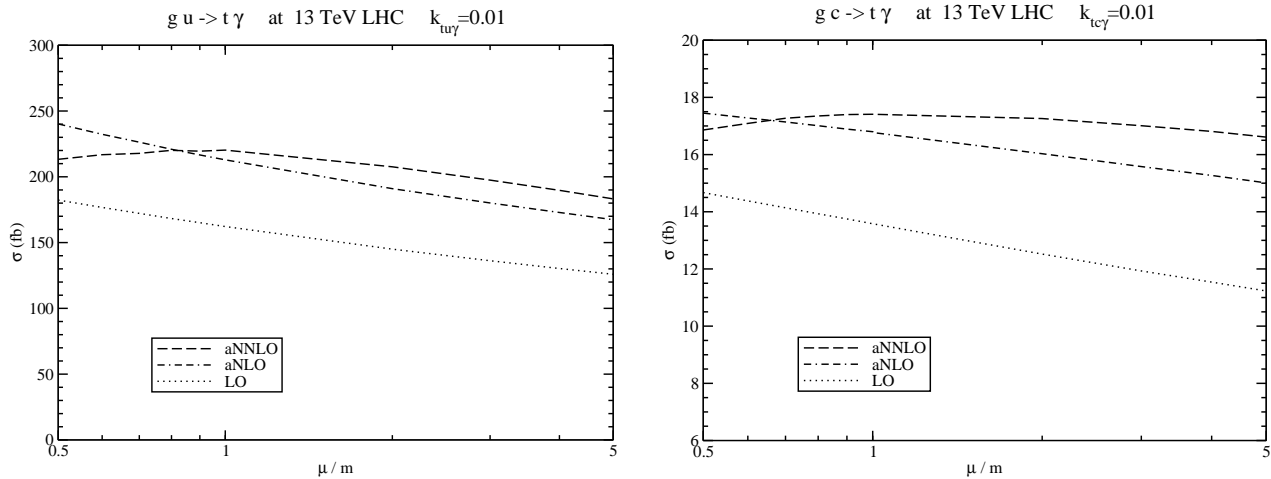

\begin{center}
\includegraphics[width=81mm]{gutgam13mu.eps}
\hspace{3mm}
\includegraphics[width=81mm]{gctgam13mu.eps}
\caption{The scale dependence of the total cross sections for $t\gamma$ production via the processes $gu\rightarrow t\gamma$ (left) and $gc\rightarrow t\gamma$ (right) at 13 TeV LHC energy with $m=173$ GeV.}
\label{gutgammu}
\end{center}
\end{figure}

We now look at the scale dependence of the cross sections. We choose to show results using the current LHC energy of 13 TeV and a top-quark mass of 173 GeV. Fig. \ref{gutgammu} displays numerical results for the cross section as a function of scale in the range of $\mu = 0.5m$ to $\mu = 5m$, where $m$ is the top quark mass. The left plot shows results for the process $gu \rightarrow t\gamma$, and the right plot shows results for the process $gc \rightarrow t\gamma$. The percentage variation between the largest and smallest values in the range, relative to their average, is significantly different at different orders.

For the process $gu \rightarrow t\gamma$ at 13 TeV energy, we find an aNNLO scale variation of 18.4\% over the range of $\mu = 0.5m$ to $\mu = 5m$, which is significantly smaller than the aNLO variation of 35.7\%. For other LHC energies, we find aNNLO scale variations of 17.6\% at 14 TeV, 24.2\% at 8 TeV, and 26.1\% at 7 TeV. While we chose a large range for the variation, we note that a scale variation by a factor of two is typically used to estimate theoretical uncertainty. In the typical range of $\mu = 0.5m$ to $\mu = 2m$, we find aNNLO scale variations between maximum and minimum values of 5.4\% at 14 TeV, 5.9\% at 13 TeV, 11.6\% at 8 TeV, and 13.5\% at 7 TeV.

For the process $gc \rightarrow t\gamma$ at 13 TeV, we find an aNNLO scale variation of 4.7\%, over the range of $\mu = 0.5m$ to $\mu = 5m$, which is much smaller than the aNLO variation of 15.0\%. We find similar aNNLO scale variations over this range at other LHC energies: 4.5\% at 14 TeV, 7.0\% at 8 TeV, and 8.2\% at 7 TeV. In the typical range of $\mu = 0.5m$ to $\mu = 2m$, we find aNNLO scale variations of 3.9\% at 14 TeV, 3.2\% at 13 TeV, 2.6\% at 8 TeV, and 3.7\% at 7 TeV.

When considering theoretical uncertainties, pdf uncertainties must also be included. We find pdf uncertainties to be fairly small for the process $gu \rightarrow t\gamma$, but not so small for the process $gc \rightarrow t\gamma$, using the pdf errors associated with MMHT2014 \cite{MMHT} NNLO pdfs. For $gu \rightarrow t\gamma$, we find uncertainties of 1.1\% at 14 TeV, 1.1\% at 13 TeV, 1.0\% at 8 TeV, and 1.1\% at 7 TeV. For $gc \rightarrow t\gamma$, we find uncertainties of 2.7\% at 14 TeV, 2.8\% at 13 TeV, 3.6\% at 8 TeV, and 3.9\% at 7 TeV.

In principle there are uncertainties from other terms beyond the ones included in our formulas at NLL accuracy. Of course we do not know all those terms, but we know some of them, such as the two-loop anomalous dimension, Eq. (\ref{tgam2l}). If we include this two-loop term in the aNNLO expansion, we find that its effect is less than one per mille for the cross section, which is negligible when contrasted to the scale and pdf uncertainties.

\begin{figure}
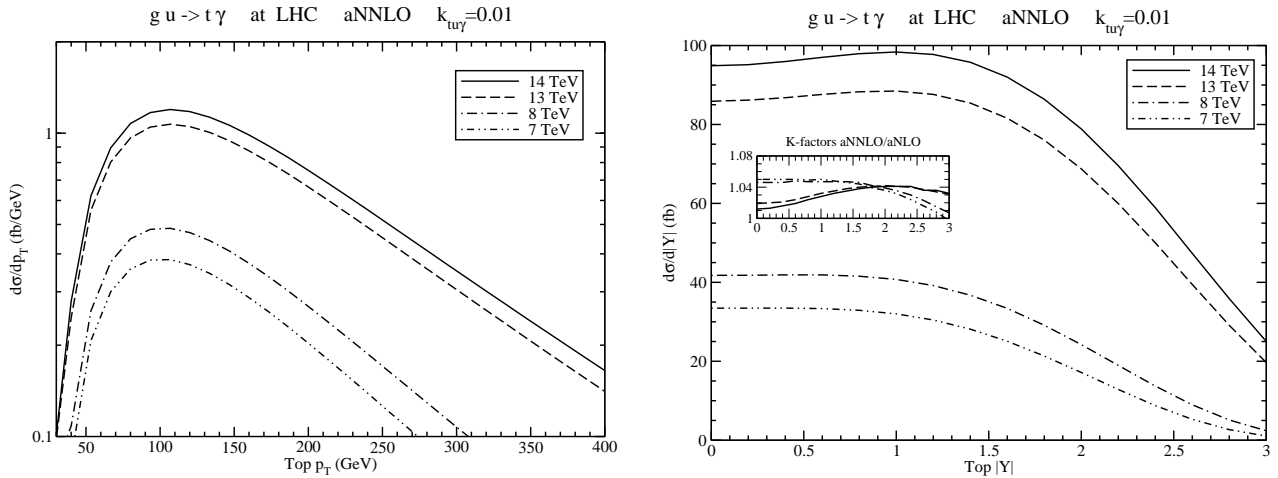

\begin{center}
\includegraphics[width=81mm]{gutgamNNLOpt.eps}
\hspace{3mm}
\includegraphics[width=81mm]{gutgamNNLOY.eps}
\caption{The top-quark $p_T$ (left) and rapidity (right) distributions in the process $gu\rightarrow t\gamma$ at LHC energies with $m=173$ GeV.}
\label{ptygutgam}
\end{center}
\end{figure}

We now consider the top-quark transverse momentum ($p_{T}$) and rapidity distributions up to aNNLO at LHC energies. The renormalization and factorization scales are again set equal to the top quark mass, chosen to be 173 GeV once more.

For the process $gu \rightarrow t\gamma$, we present aNNLO top-quark $p_T$ distributions ($d\sigma/dp_T$) and top-quark rapidity distributions ($d\sigma/d|Y|$) in the left and right plots of Fig. \ref{ptygutgam}, respectively. The distributions are shown at LHC energies of 7, 8, 13, and 14 TeV, once again. The $p_T$ distributions peak at a $p_T$ value of about 110 GeV for LHC energies of 13 and 14 TeV and about 100 GeV for energies of 7 and 8 TeV.

\begin{figure}
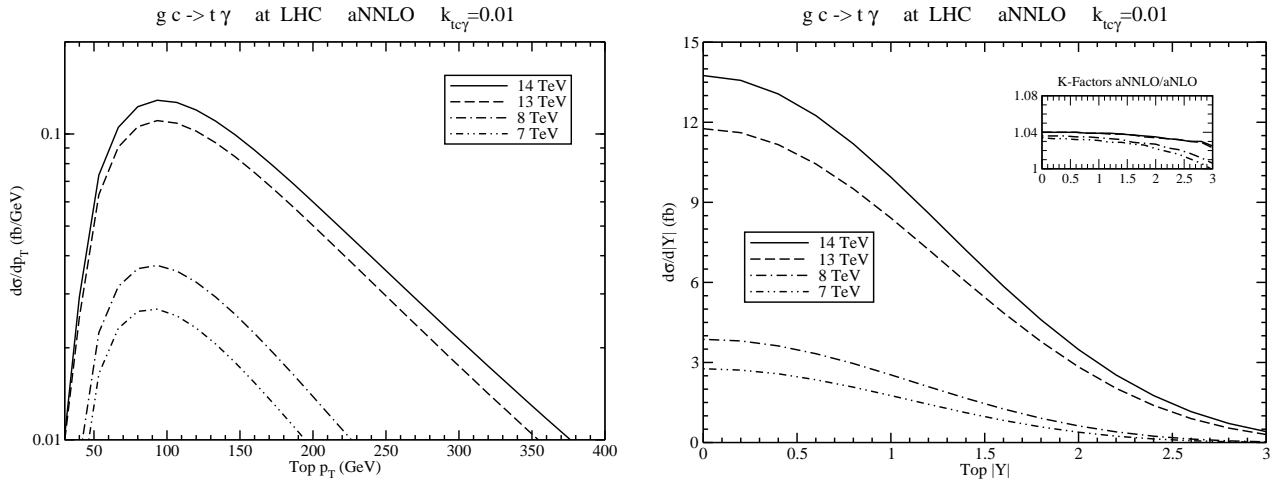

\begin{center}
\includegraphics[width=81mm]{gctgamNNLOpt.eps}
\hspace{3mm}
\includegraphics[width=81mm]{gctgamNNLOY.eps}
\caption{The top-quark $p_T$ (left) and rapidity (right) distributions in the process $gc\rightarrow t\gamma$ at LHC energies with $m=173$ GeV.}
\label{ptygctgam}
\end{center}
\end{figure}

Finally, in Fig. \ref{ptygctgam} we present the aNNLO top-quark $p_T$ and rapidity distributions for the process $gc \rightarrow t\gamma$. The magnitudes of both the rapidity and $p_T$ distributions are once again much smaller than those for the process $gu \rightarrow t\gamma$, just as with the total cross sections. The $p_T$ distributions peak at a $p_T$ value of about 100 TeV for LHC energies of 13 and 14 TeV and about 90 GeV for energies of 7 and 8 TeV. 

\mysection{Conclusions}

We have studied the production of a top quark in association with a photon via the partonic processes $gu \rightarrow t\gamma$ and $gc \rightarrow t\gamma$, which involve anomalous $t$-$q$-$\gamma$ couplings. We have found large contributions from soft-gluon emission for the total production cross sections and the top-quark transverse-momentum and rapidity distributions. The soft-gluon contributions dominate the cross section numerically, and in fact at NLO they approximate very well the complete results. We have resummed soft-gluon emission for these processes and determined the relevant soft anomalous dimensions through two loops. From the NLL resummed cross section we have derived expansions through NNLO. The aNNLO soft-gluon corrections provide substantial additional enhancements.

The total cross sections and the top-quark transverse-momentum and rapidity distributions were calculated for the $gu \rightarrow t\gamma$ and $gc \rightarrow t\gamma$ processes at 7, 8, 13, and 14 TeV LHC energies. The aNNLO soft-gluon corrections are significant at all LHC energies. Hence, these corrections must be included for improved theoretical input in searching for processes with anomalous couplings and in setting limits for those couplings. 

\mysection*{Acknowledgements}
This material is based upon work supported by the National Science Foundation under Grant No. PHY 1519606 and PHY 1820795.


\begin{thebibliography}{99}

\bibitem{NKrev}
N. Kidonakis, Int. J. Mod. Phys. A {\bf 33}, 1830021 (2018) [arXiv:1806.03336 [hep-ph]].

\bibitem{TY}
T. Tait and C.-P. Yuan, Phys. Rev. D {\bf 63}, 014018 (2000) [hep-ph/0007298].

\bibitem{NKAB}
N. Kidonakis and A. Belyaev, JHEP {\bf 0312}, 004 (2003) [hep-ph/0310299]. 

\bibitem{ZLLGZ}
Y. Zhang, B.H. Li, C.S. Li, J. Gao, and H.X. Zhu, Phys. Rev. D {\bf 83}, 094003 (2011) [arXiv:1101.5346 [hep-ph]]. 

\bibitem{FG}
M. Fael and T. Gehrmann, Phys. Rev. D {\bf 88}, 033003 (2013) 
[arXiv:1307.1349 [hep-ph]].

\bibitem{TopC}
J. Adelman {\it et al.}, arXiv:1309.1947 [hep-ex].

\bibitem{DMWZ}
C. Degrande, F. Maltoni, J. Wang, and C. Zhang, Phys. Rev. D {\bf 91}, 034024 (2015) [arXiv:1412.5594 [hep-ph]].

\bibitem{DMZ}
G. Durieux, F. Maltoni, and C. Zhang, Phys. Rev. D {\bf 91}, 074017 (2015) [arXiv:1412.7166 [hep-ph]]. 

\bibitem{GYY}
Y.-C. Guo, C.-X. Yue, and S. Yang, Eur. Phys. J. C {\bf 76}, 596 (2016) [arXiv:1603.00604 [hep-ph]].

\bibitem{CMS}
CMS Collaboration, JHEP {\bf 1604}, 035 (2016) [arXiv:1511.03951 [hep-ex]].

\bibitem{NKtop}	
N. Kidonakis, Phys. Rev. D {\bf 90}, 014006 (2014) [arXiv:1405.7046 [hep-ph]];
{\bf 91}, 031501(R) (2015) [arXiv:1411.2633 [hep-ph]]; 
{\bf 91}, 071502(R) (2015) [arXiv:1501.01581 [hep-ph]];
in Proceedings of HQ2013 [arXiv:1311.0283 [hep-ph]].

\bibitem{NKts}
N. Kidonakis, Phys. Rev. D {\bf 81}, 054028 (2010) [arXiv:1001.5034 [hep-ph];
{\bf 83}, 091503(R) (2011) [arXiv:1103.2792 [hep-ph]].

\bibitem{NKtW}
N. Kidonakis, Phys. Rev. D {\bf 82}, 054018 (2010) [arXiv:1005.4451 [hep-ph]];
{\bf 96}, 034014 (2017) [arXiv:1612.06426 [hep-ph]].

\bibitem{NKtZ}
N. Kidonakis, Phys. Rev. D {\bf 97}, 034028 (2018) [arXiv:1712.01144 [hep-ph]].

\bibitem{NKGS}
N. Kidonakis and G. Sterman,  Nucl. Phys. B {\bf 505}, 321 (1997) [hep-ph/9705234].

\bibitem{NK2loop}
N. Kidonakis, Phys. Rev. Lett. {\bf 102}, 232003 (2009) [arXiv:0903.2561 [hep-ph]]; Int. J. Mod. Phys. A {\bf 31}, 1650076 (2016) [arXiv:1601.01666 [hep-ph]]. 

\bibitem{GS87}
G. Sterman, Nucl. Phys. B {\bf 281}, 310 (1987). 

\bibitem{CT89}
S. Catani and L. Trentadue, Nucl. Phys. B {\bf 327}, 323 (1989).

\bibitem{MMHT}
L.A. Harland-Lang, A.D. Martin, P. Molytinski, and R.S. Thorne,   
Eur. Phys. J. C {\bf 75}, 204 (2015) [arXiv:1412.3989 [hep-ph]].

\bibitem{CT14}
S. Dulat, T.-J. Hou, J. Gao, M. Guzzi, J. Huston, P. Nadolsky, J. Pumplin, 
C. Schmidt, D. Stump, and C.-P. Yuan, Phys. Rev. D {\bf 93}, 033006 (2016) 
[arXiv:1506.07443 [hep-ph]].

\bibitem{NNPDF}
NNPDF Collaboration, R.D. Ball {\it et al.}, 
Eur. Phys. J. C {\bf 77}, 663 (2017) [arXiv:1706.00428 [hep-ph]]. 

\end{thebibliography}
\end{document}